\documentclass{article}

\usepackage[utf8x]{inputenc}
\usepackage[T1]{fontenc}
\usepackage{geometry}
\usepackage{graphicx}
\usepackage{svg}

\usepackage[sort&compress]{natbib}

\setcitestyle{square}
\bibpunct{[}{]}{,}{n}{,}{,}
\usepackage{caption}
\usepackage{subcaption}
\usepackage{mathtools}
\usepackage{textgreek}
\usepackage{amsmath}
\usepackage{authblk}

\usepackage{amssymb}
\usepackage{float}
\usepackage{upgreek}
\usepackage{bbm}
\usepackage{xcolor}

\usepackage[pdftex,colorlinks=true,pdfstartview=Fit]{hyperref}
\usepackage{physics}
\DeclareMathOperator{\e}{e}
\newcommand{\sub}[1]{_{\mbox{\scriptsize{#1}}}}  

\newcommand{\tn}{\textnormal}

\title{Performance limits of information engines}

\author[1]{Johan du Buisson\thanks{johannes\_du\_buisson@sfu.ca}}
\author[1$\&$]{David A.\ Sivak\thanks{dsivak@sfu.ca}}
\author[1$\&$]{John Bechhoefer\thanks{johnb@sfu.ca}}
\affil[1]{Dept.\ of Physics, Simon Fraser University}
\affil[$\&$]{These authors contributed equally.}
\date{}

\begin{document}

\maketitle

\begin{abstract}
We review recent studies of a colloidal information engine that consists of a bead in water and held by an optical trap. The bead is ratcheted upward without any apparent external work, by taking advantage of favorable thermal fluctuations. Much of the previous work on such engines aimed to show that accounting for information-processing costs can reconcile the observed motion with the second law of thermodynamics. By contrast, we focus on the factors that limit the performance of such engines by optimizing variously the upward velocity, rate of gravitational free-energy extraction, or ability to track a trajectory. We then consider measurement noise, which degrades engine performance. A naive use of noisy measurements in the feedback algorithm leads to a phase transition at finite signal-to-noise ratio:  below the transition, the engine no longer functions. A more sophisticated, `Bayesian' algorithm eliminates the phase transition and improves performance. Finally, operating the information engine in a nonequilibrium environment with extra force fluctuations can enhance the performance by orders of magnitude, even to the point where the energy extracted exceeds that needed to run the information processing. Autonomous implementations of an information engine in such environments could be powered entirely by the additional energy of the bath.

\end{abstract}

\section{Introduction}
Over 150 years ago, Maxwell~\cite{maxwell1871theory} proposed the following thought experiment: a volume of gas is separated by a partition into two chambers, A and B. A `neat fingered being' (later referred to as a `demon') which has knowledge of the velocities of the gas molecules allows only fast molecules to pass through an opening in the partition from A to B, and only slow molecules to pass from B to A, creating a temperature difference between the chambers, which can be used to do work. This sorting reduces the entropy of the gas, without any work being done directly on the system, in apparent violation of the second law of thermodynamics. 

In 1929, Szilard~\cite{szilard1964decrease} proposed a simpler version of this thought experiment, where a single gas molecule moves in a chamber whose walls are maintained at a temperature $T$ and which is divided by a partition. If the molecule is observed in the left half of the chamber, a mass is attached that is raised by motion to the right, and vice versa. An isothermal expansion of the gas then raises the mass, storing energy via the gravitational potential. Cyclic operation of this engine uses information regarding the molecule's position (whether the molecule is located in the left or right half of the container) to convert heat into stored gravitational energy, in seeming contradiction of the second law. These thought experiments inspired the notion of an \textit{information engine} and led to deeper investigations into the role that information plays in thermodynamic processes and the physical nature and thermodynamics of information~\cite{landauer1991information, Parrondo2015_Thermodynamics}. 

To function, a Szilard engine stores a molecule's position as a bit of memory (0 when left of the barrier, 1 when right), which must be subsequently erased during cyclic operation. As argued by Landauer~\cite{landauer1961irreversibility}, the corresponding reduction in entropy requires a minimum energy of $k_{\tn{B}} T \ln 2$. This observation, now known as \textit{Landauer's principle}, resolves the apparent violation of the second law. 

The last few decades have seen remarkable progress in our understanding of thermodynamics, driven by the development of stochastic thermodynamics~\cite{sekimoto1997kinetic,sekimoto2010stochastic,seifert2012stochastic}, which extends the notions of classical thermodynamics (such as heat, work and entropy) to individual trajectories of mesoscopic systems in contact with a heat bath. This framework has produced powerful novel results, most notably fluctuation theorems~\cite{gallavotti1995dynamical2, gallavotti1995dynamical,jarzynski1997nonequilibrium,crooks1999entropy,kurchan1998fluctuation,maes1999fluctuation}, which constrain the fluctuations of these thermodynamic quantities in and out of equilibrium and refine the second law of thermodynamics. Sagawa~\cite{sagawa2008second,sagawa2010generalized} and others~\cite{cao2009thermodynamics} extended these results to feedback-controlled systems, showing how to modify thermodynamic relationships to account for the costs of information processing. This theoretical work 
confirmed that Landauer's principle is a consequence of the second law of thermodynamics when including the entropy changes associated with information gain or manipulation. 

The theoretical insights offered by stochastic thermodynamics coupled with technological breakthroughs in the observation and control of small systems prompted Toyabe, et al.~\cite{toyabe2010experimental} to physically realize a Maxwell demon and explicitly demonstrate information-to-work conversion. They showed that a Brownian particle under nonequilibrium feedback control can climb a `spiral-staircase potential', converting particle-position information into free energy in excess of the work directly performed on it by the feedback apparatus. This work also experimentally verified the generalized fluctuation relation obtained by Sagawa. Since then, researchers have realized a variety of information engines~\cite{camati2016experimental,koski2014experimental2,koski2015chip,cottet2017observing,masuyama2018information,chida2017power,Paneru2018_Losless,paneru2018optimal,paneru2020efficiency,Admon2018_Experimental,price2008single,kumar2018sorting,raizen2009comprehensive,thorn2008experimental}, concrete implementations of the thought experiments of Maxwell and Szilard.

Realizations more directly analogous to Maxwell's original concept that use information to sort atoms have also been realized~\cite{price2008single,kumar2018sorting,raizen2009comprehensive,thorn2008experimental}. In pioneering work, Thorn, et al.~\cite{thorn2008experimental} implemented a Maxwell demon that separated atoms in their ground state from those in a higher-energy state. Later, Pekola and coworkers used a single-electron box to demonstrate the extraction of $k_{\tn{B}} T \ln 2$ of heat from the bath per bit of information (encoded by the position of an electron)~\cite{koski2014experimental2}. These Maxwell demons were all external to the system acted upon (the feedback is applied by an external agent), complicating the analysis of the thermodynamics. By capacitively linking two single-electron devices in one electronic circuit~\cite{koski2015chip}, the Pekola group constructed an autonomous Maxwell demon (operating without external control), where the demon and system together constitute a single super-system that can be studied holistically, allowing explicit experimental observation of the thermodynamics associated with the demon's information processing. 

The ability to apply feedback to small systems and to physically realize information engines has also been used to experimentally study the cost of information processing and to provide further support for Landauer's principle~\cite{berut2012experimental,jun2014high, koski2014experimental,hong2016experimental,ribezzi2019large,sairi2020nonequilibrium,wimsatt2021harnessing}, including extensions to finite-time bit erasure~\cite{scandi2022minimally} and underdamped systems~\cite{dago2021information,dago2022dynamics,dago2023adiabatic}, while Admon, et al.~\cite{Admon2018_Experimental} and Paneru, et al.~\cite{paneru2018optimal,Paneru2018_Losless,paneru2020efficiency} have quantified the efficiency of information-to-work conversion in feedback-controlled colloidal particles~\cite{paneru2020colloidal}. Feedback-controlled systems that behave similarly to information engines have been used to cool nanoparticles to millikelvin temperatures~\cite{li2012fundamental,gieseler2012subkelvin,tebbenjohanns2019cold}, and the ideas of Maxwell, Szilard and Landauer have inspired the construction of molecular information engines~\cite{serreli2007molecular}, which shows that synthetic molecular machines can function as information engines.

Here, we review recent studies~\cite{Saha2021_Maximizing,lucero2021maximal,saha2021trajectory,saha2022bayesian,saha2023information} of performance limits of an information engine that extracts energy from a heat bath and stores this energy by raising a weight, as initially imagined by Szilard. Rather than focusing on the question of compatibility with the second law, we instead ask how well such engines can function. 

Our information engine consists of an optically trapped, micron-scale bead in water, with the trap's horizontal laser beam perpendicular to the vertical gravitational axis. The optical tweezers create a harmonic potential, while gravity causes the bead to sag, so that its minimum-energy position lies below the trap center. The trap is shifted upwards whenever the bead's position fluctuates above a particular value. This raises the bead's average equilibrium position, thereby storing gravitational energy. The experimental setup is similar to earlier colloidal information engines~\cite{lopez2008realization,Paneru2018_Losless}. A novel feature of the feedback rule employed is that it allows for a `pure' information ratchet: the trap center (owing to the nature of the quadratic potential) can be shifted without doing external work on the bead, simplifying the energetics and ensuring that the only cost is the energy needed to operate the measurement-and-feedback apparatus~\cite{ehrich2023energetic}. 

\begin{figure}[t!]
    \centering
    \includegraphics[width = 0.7\linewidth]{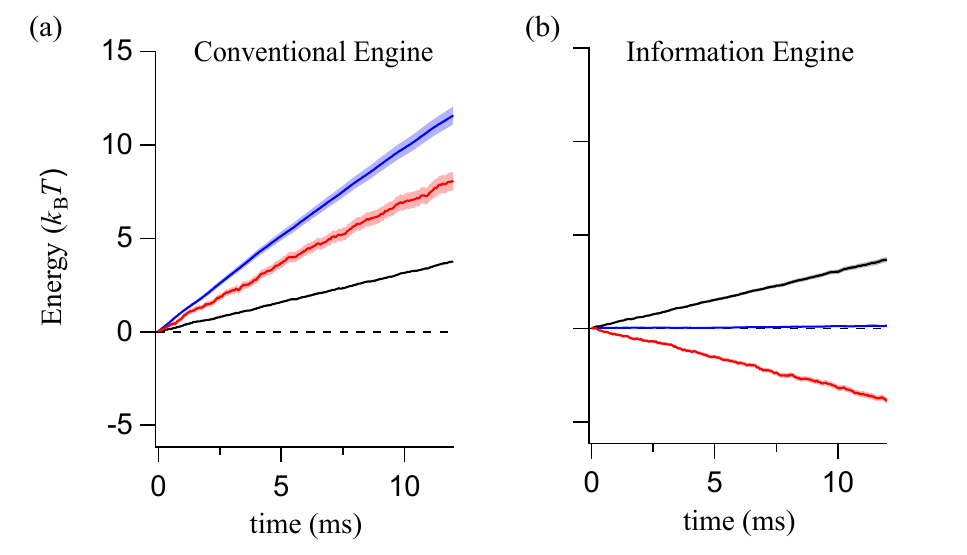}
    \caption{Energy flows in conventional and information engines used to displace a bead. (a) In a conventional engine, the mean work (blue) equals the sum of dissipated heat (red) and change in bead free energy (black). (b) In a pure information ratchet, the work applied is zero (blue), so that the extracted power (black) is supplied by heat \textit{extracted} from the bath (red).  In the two engines, the parameters are adjusted to match the engine power output (black). Adapted from~\cite{saha2022PhD}.}
    \label{fig:infoVconventional}
\end{figure}

Before presenting our system in detail, we illustrate how an information engine differs from conventional engines, using data from our setup~\cite{saha2022PhD}. Figure~\ref{fig:infoVconventional} contrasts energy flows for a conventional engine (a), where the bead is dragged up by a constant force through a fluid, and an information ratchet (b), where the bead is lifted through the action of the information engine described above. From the first law of thermodynamics, the work equals the increase in the gravitational free energy of the bead plus the heat dissipated by the movement of the bead through the viscous fluid. Gravitational potential energy increases at the same rate in both cases; however, in (a) the work done on the bead is partially dissipated as heat in the bath, whereas in (b) the work applied to the bead through the trap is zero. The energy to raise the bead now comes from heat extracted from the bath.  Thus, in (a), the motion of the bead \textit{heats} the bath; in (b), it \textit{cools} the bath.

We study and quantify the performance of this engine, posing several questions: What are the maximum directed velocity and gravitational energy storage~\cite{Saha2021_Maximizing}? What feedback strategies maximize net power output when we allow positive trap work~\cite{lucero2021maximal}? What signal bandwidth can we track when we not only rectify motion but control the bead's overall trajectory~\cite{saha2021trajectory}?  How does noise from the measurement of the bead degrade performance~\cite{saha2022bayesian}?  Can placing the trap in a nonequilibrium environment with `extra' fluctuations improve performance~\cite{saha2023information}? Below, we review some answers to these questions.

\section{Materials and methods} \label{sec:materialsmethods}
The dynamics of a harmonically trapped, micron-scale bead in water are well-described by the overdamped Langevin equation 
\begin{equation} \label{langevin_main}
    \gamma \dot{x}(t) = \underbrace{- \kappa [x(t) - \lambda(t)]
    }_{\tn{restoring force}} \,\,- \underbrace{mg}_{\tn{grav. force}} + \, \, \underbrace{\sqrt{2 k_{\tn{B}} T \gamma} \,\, \xi(t)}_{\tn{thermal noise}}, 
\end{equation}
for bead position $x(t)$, trap center $\lambda(t)$, trap stiffness $\kappa$, bead friction coefficient $\gamma$, net mass $m = \Delta \rho V$ (with $\Delta \rho$ the density difference between bead and fluid and $V$ the bead volume), gravitational acceleration $g$, temperature $T$, and Gaussian white noise $\xi(t)$ with zero mean and temporal autocorrelation $\langle \xi(t) \, \xi(t') \rangle = \delta(t - t')$. 

Scaling lengths by the equilibrium standard deviation $\sigma = \sqrt{k_{\tn{B}} T /\kappa}$ of the bead position and time by the bead relaxation time $\tau_{\tn{r}} = \gamma/\kappa$, the Langevin equation becomes
\begin{equation} 
\label{langevin_scaled}
    \dot{x}(t) = - (x(t) - \lambda(t)) - \delta_{\tn{g}} + \sqrt{2} \, \xi(t),
\end{equation}
with $\delta_{\tn{g}} \equiv mg/\kappa \sigma$ the bead's scaled effective mass, which accounts for the effects of gravity and buoyancy. 

The bead position $x_n$ is measured at a sampling interval $t_{\tn{s}}' = 20$ \textmu s, following which the position of the trap center $\lambda_{n+1}$ is updated. The feedback has a delay of one time step. As such, integrating (\ref{langevin_scaled}) over one time step~\cite{kloeden1992stochastic} gives the stochastic discrete-time update rule for the bead position,
\begin{equation}
    x_{n + 1} = \e^{-t_{\tn{s}}} x_n + (1 - \e^{-t_{\tn{s}}})(\lambda_n - \delta_{\tn{g}}) + \sqrt{1 - \e^{-2 t_{\tn{s}}}} \, \xi_n,
\end{equation}
with $t\sub{s} = t\sub{s}'/\tau\sub{r}$ and $\xi_n$ an i.i.d. Gaussian random variable with zero mean and covariance $\langle \xi_n \xi_m \rangle = \delta_{nm}$. 

\begin{figure}
    \centering
    \includegraphics[width=\linewidth]{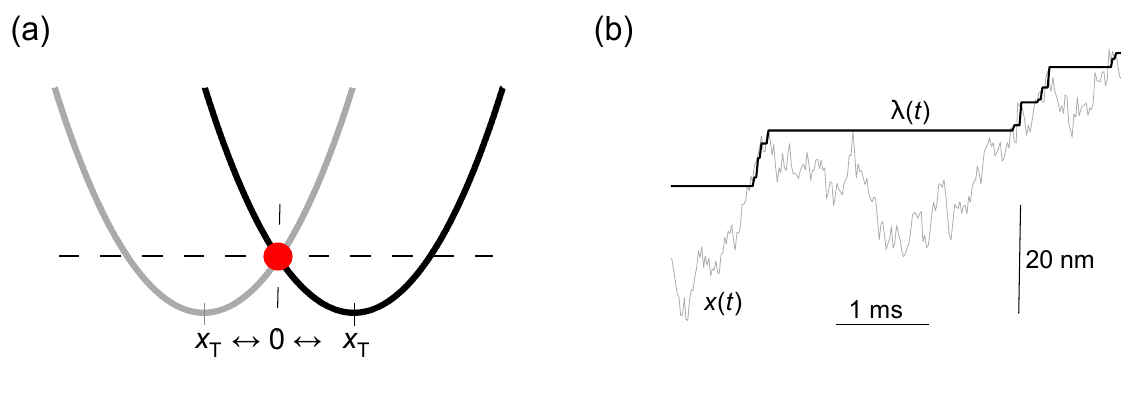}
    \caption{The feedback rule and a representative trajectory for a pure information engine. (a) The harmonic potential is shown before (gray) and after (black) trap update, with the bead position indicated in red. The naive zero-work condition corresponds to the case $\alpha = 2$, for which the bead is located a distance $X_{\tn{T}}$ from the trap center before and after the ratchet event. (b) Experimental trajectory of the bead (gray) and trap (black). Adapted from~\cite{Saha2021_Maximizing}.}
    \label{fig:pnasFigure2}
\end{figure}

The trap position is updated according to the feedback rule
\begin{equation} \label{update_main}
    \lambda_{n+1} = \lambda_n + \alpha (x_n - \lambda_n) \, \Theta(x_n - \lambda_n - X_{\tn{T}}),
\end{equation}
for feedback gain $\alpha$ (a positive constant), Heaviside step function $\Theta(\cdot)$, and threshold position $X_{\tn{T}}$. In words, the trap remains stationary until fluctuations drive $x - \lambda$ above $X_{\tn{T}}$, at which point the trap `ratchets', moving up by an amount set by the value of the feedback gain constant. This results in an increase in the bead's average position and consequently in the storage of gravitational potential energy. The threshold $X_{\tn{T}}$ sets the minimum fluctuation that is captured and rectified. Figure~\ref{fig:pnasFigure2}(a) shows the relative positions of bead and trap before and after a ratchet event, and Fig.~\ref{fig:pnasFigure2}(b) illustrates a representative trajectory.

The trap is stationary for duration $t_{\tn{s}}$ following measurement of the bead position, and the trap center $\lambda$ changes much faster than the bead position $x$, so the input work to shift the trap center at time $t_{n+1}$ is approximately
\begin{equation} 
\label{trap_work}
    W_{n + 1} = \frac{1}{2} \left[(x_{n + 1} - \lambda_{n + 1})^2 - (x_{n + 1} - \lambda_n)^2 \right] \,.
\end{equation}
We store gravitational energy solely by exploiting information regarding the bead position in order to rectify beneficial `up' fluctuations, with no trap work. This `pure' information engine is distinct from other Maxwell-demon-type systems that combine information and external work to perform useful functions. The zero-trap-work requirement amounts, in the limit where the measurement and subsequent trap update happen instantaneously, to the choice $\alpha = 2$. However, the delay $t_{\tn{s}}$ between the position measurement and the feedback requires that a smaller $\alpha$ be used to account for the bead's tendency to move down towards its average equilibrium position due to gravity. 

Finally, the gravitational free energy stored during a trap update is 
\begin{equation}
    \Delta F_{n + 1} = \delta_{\tn{g}} (\lambda_{n + 1} - \lambda_n) \,. 
\end{equation}
Note that no violation of the second law occurs, even though apparently zero work is applied to store gravitational free energy: The necessary storage, processing, and erasure of information comes with fundamental thermodynamic costs~\cite{Parrondo2015_Thermodynamics}, i.e., a nonzero minimum work to apply feedback control~\cite{ehrich2023energetic}. When the system and measuring apparatus are at the same temperature, the power extracted as stored gravitational free energy cannot exceed the costs of running the engine.

\subsection{Experimental apparatus}

\begin{figure}[h]
    \centering
    \includegraphics[width=0.7\linewidth]{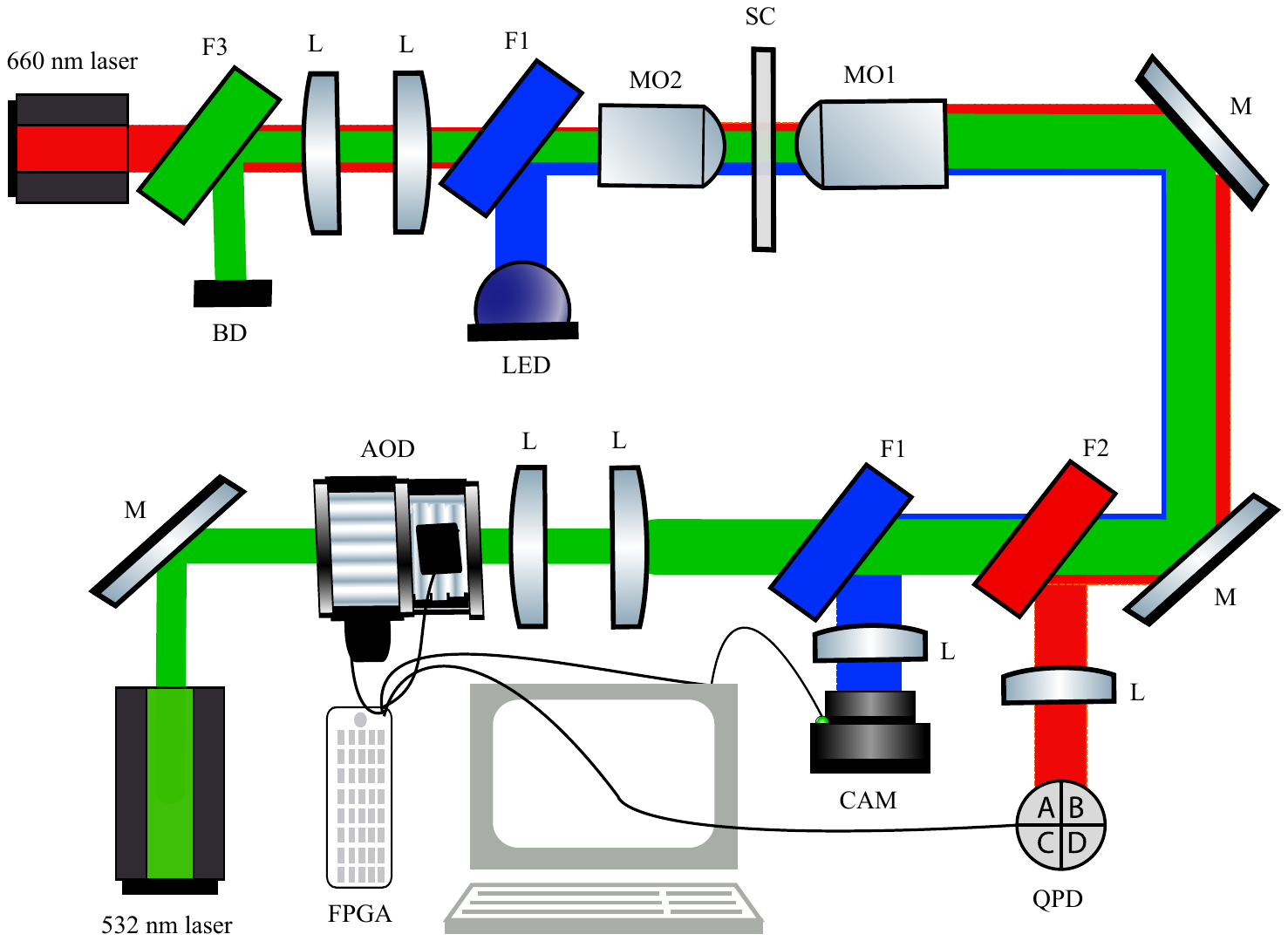}
    \caption{Schematic diagram of the experimental set-up. AOD = Acousto-Optic Deflector, FPGA $=$ Field-Programmable Gate Array, QPD = Quadrant Photodiode, L = lenses, M = mirrors, F1--3 = dichroic mirrors. Representative components are not to scale. From~\cite{Saha2021_Maximizing}.}
    \label{fig:experimental_setup}
\end{figure}

Figure~\ref{fig:experimental_setup} shows essential parts of the experimental set-up~\cite{kumar2018nanoscale,Saha2021_Maximizing}. A micron-scale silica bead is trapped by optical tweezers whose position can be rapidly adjusted. Optical trapping is produced by a tightly focused 532-nm green laser using a microscope objective (60x water immersion, NA = 1.2). A 660-nm red detection laser is loosely focused on the bead using a second objective (40x, NA = 0.5). The two objectives are aligned to match their focal planes inside the sample chamber. Two acousto-optic deflectors (AODs) steer, at microsecond time scales, the position of the trapping beam parallel and perpendicular to gravity. The red (660-nm) light collected from the trapping plane is focused on a quadrant photodiode for position detection. The bead position is measured every 20 \textmu s (50 kHz) by the A/D converter of a Field Programmable Gate Array, which calculates the feedback signal for the AOD.

\section{Maximizing directed velocity and power extraction} 
\label{sec:max_vel_power}

Here we present results from \cite{Saha2021_Maximizing} on the theoretical limits for performance of our information engine while disregarding the costs associated with processing information and applying feedback. First, we study the fundamental upper bounds on the directed velocity and power extraction for zero-trap-work feedback. How fast can it go? How much can it lift? Understanding these output limits is important for two reasons: First, in practical scenarios the benefit of performing some function often far outweighs the energetic cost; e.g., in biological applications such as chemotaxis, the metabolic costs of operating signal-processing machinery are usually unimportant compared to the benefit gained from acquiring a new food source or escaping a predator~\cite{berg2004coli}. Second, these upper limits serve as performance benchmarks when also considering trade-offs with operating costs.

Throughout this section, we assume that measurements are noise-free; this is a good approximation when such errors are much smaller than the characteristic lengthscale $\sigma$, the bead's equilibrium standard deviation. 

The average directed velocity $v$ is the change in the trap center divided by the elapsed time for an infinitely long trajectory of the dynamics (Eqs.~\ref{langevin_scaled} and \ref{update_main}):
\begin{equation}
    v = \lim_{t \rightarrow \infty} \frac{\lambda(t) - \lambda(0)}{t} \,.
\end{equation}
The associated average rate of free-energy extraction (or output power) is, in scaled units,
\begin{equation}
    \dot{F} = \delta\sub{g} v \ .
\end{equation}

\subsection{Effect of sampling frequency}
\begin{figure}
    \centering
    \includegraphics[width = \linewidth]{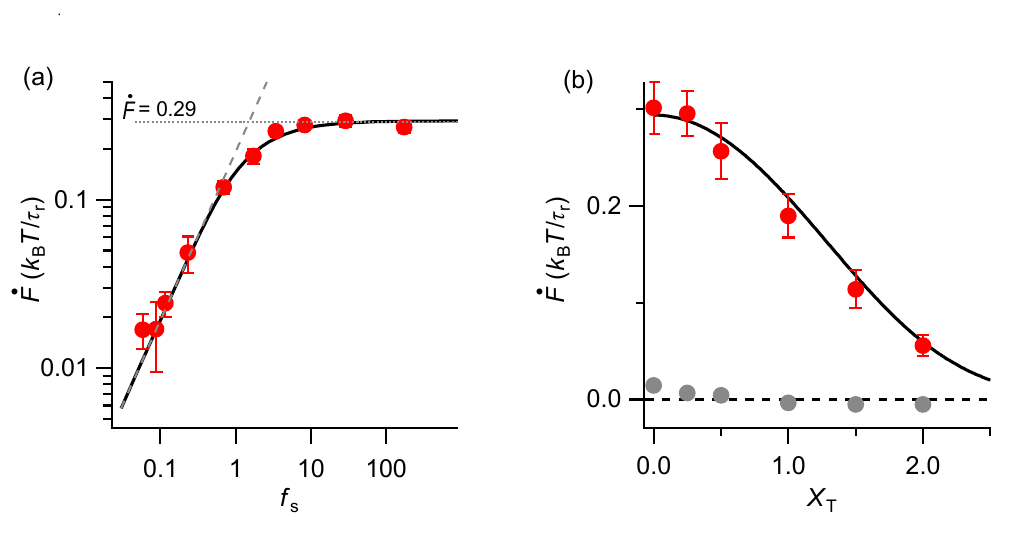}
    \caption{Optimization of output power. (a) Output power as a function of sampling frequency $f_s = \tau\sub{r}/t\sub{s}'$, with semi-analytic results (black) compared with experimental values (red) for $X\sub{T} = 0$ and showing the infinite-frequency limit (gray, dotted) and the low-frequency limit (gray, dashed). (b) Power as a function of threshold value $X\sub{T}$ for fixed feedback gain and sampling frequency. The trap input power (gray markers) is small, and analytic calculations (black curve) show good agreement with experiment (red). Adapted from~\cite{Saha2021_Maximizing}.}
    \label{fig:pnasFigure3}
\end{figure}

Here we investigate how information-engine performance varies with sampling frequency $f_{\tn{s}} = \tau_{\tn{r}}/t_{\tn{s}}$, at fixed trap stiffness $\kappa$. Figure~\ref{fig:pnasFigure3}(a) shows that the extracted power increases monotonically as the sampling frequency increases, increasing linearly for small $f_{\tn{s}}$ and saturating at high $f_{\tn{s}}$. For low sampling frequencies, the information engine misses beneficial `up' fluctuations, so that increasing $f_{\tn{s}}$ significantly increases the extracted power. As $f_{\tn{s}}$ increases, a larger proportion of beneficial fluctuations are exploited, so that further increasing sampling frequency gives diminishing returns. In the limit of continuous sampling ($f_{\tn{s}} \rightarrow \infty$), all beneficial fluctuations are exploited, and performance is limited only by the rate at which such fluctuations occur, which depends on the system's material parameters and is finite. The linear scaling for low sampling frequency (where $t_{\tn{s}}$ is larger than the bead's relaxation time) results from bead equilibration between successive trap updates; the average extracted power is then the average increase in gravitational energy for a single ratchet event (from the equilibrium position distribution, hence independent of $f_{\tn{s}}$) times the sampling frequency, giving a linear dependence on $f_{\tn{s}}$.

\subsection{Optimal threshold $X_{\tn{T}}$}
Next, we consider the optimal threshold $X_{\tn{T}}$ that maximizes directed velocity and output power. We work in the continuous-sampling limit ($f_{\tn{s}} \rightarrow \infty$), where $\alpha = 2$, given that this maximizes performance. The dependence of the bead's velocity $v$ on the fluctuation threshold can be obtained by noting that following a trap update, the bead is located at $\lambda - X_{\tn{T}}$ and must reach the threshold $\lambda + X_{\tn{T}}$ before the trap ratchets up again. While the time to reach the fluctuation threshold (the first-passage time) is stochastic, the total displacement of $x$ (or $\lambda$) for a ratchet event is deterministic and equal to $2 X_{\tn{T}}$. Therefore the average directed velocity is $v = 2 X_{\tn{T}}/\tau_{\tn{MFP}}$ for the mean first-passage time $\tau_{\tn{MFP}}$. Using standard techniques~\cite{hanggi1990reaction,chupeau2020optimizing}, the mean first-passage time is
\begin{equation}
    \tau_{\tn{MFP}}(X_{\tn{T}}) = \int_{-X_{\tn{T}}}^{X_{\tn{T}}} \dd{x} \e^{V(x)} \int_{-\infty}^{x} \dd{y} \e^{-V(y)},
\end{equation}
for total potential $V(x) = \tfrac{1}{2} x^2 + \delta_{\tn{g}} x$. Intuitively, a larger $X_{\tn{T}}$ gives a larger trap update---and hence larger increase in stored gravitational energy---during any given ratchet event; however, the time to reach this higher threshold also increases. Furthermore, when the bead is above the trap, both gravity and the trap force the bead down, with the trap force increasing in strength with increasing distance from the bead to the trap. Thus, the bead must overcome a stronger downward force to fluctuate above a larger threshold $X_{\tn{T}}$. Since the probability of fluctuating above $X_{\tn{T}}$ decreases exponentially with $X_{\tn{T}}$~\cite{Saha2021_Maximizing}, the directed velocity and power are both maximized for $X_{\tn{T}} \to 0$, when individual ratchet events are small but every beneficial `up' fluctuation is exploited. This is illustrated in Fig.~\ref{fig:pnasFigure3}(b), showing that the extracted power decreases with increasing threshold $X_{\tn{T}}$.

\subsection{Optimal material parameters} \label{subsec:optimal_parameters}
As we have seen, performance saturates in the limits where the sampling time $t_{\tn{s}}$ and fluctuation threshold $X_{\tn{T}}$ tend to $0$. In these limits, and assuming a stiff trap ($\kappa \gg 1$), the maximal directed velocity (in physical units) for a given set of material parameters of the engine is found to be
\begin{equation} 
\label{vel_phys}
    v' = \left(\frac{\sigma}{\tau_{\tn{r}}} \right) v \sim \sqrt{\frac{2 k_{\tn{B}} T}{\pi}}\frac{\sqrt{\kappa}}{\gamma}
\end{equation}
with corresponding maximal output power
\begin{equation} 
\label{p_phys}
    \dot{F}' = \left(\frac{k_{\tn{B}} T}{\tau_{\tn{r}}} \right) \dot{F} \sim \sqrt{\frac{2 k_{\tn{B}} T }{\pi}}\frac{\sqrt{\kappa}}{\gamma} \, mg.
\end{equation}
The engine performance is thus limited in practice only by the system's material parameters. 
\begin{figure}
    \centering
    \includegraphics[width=\linewidth]{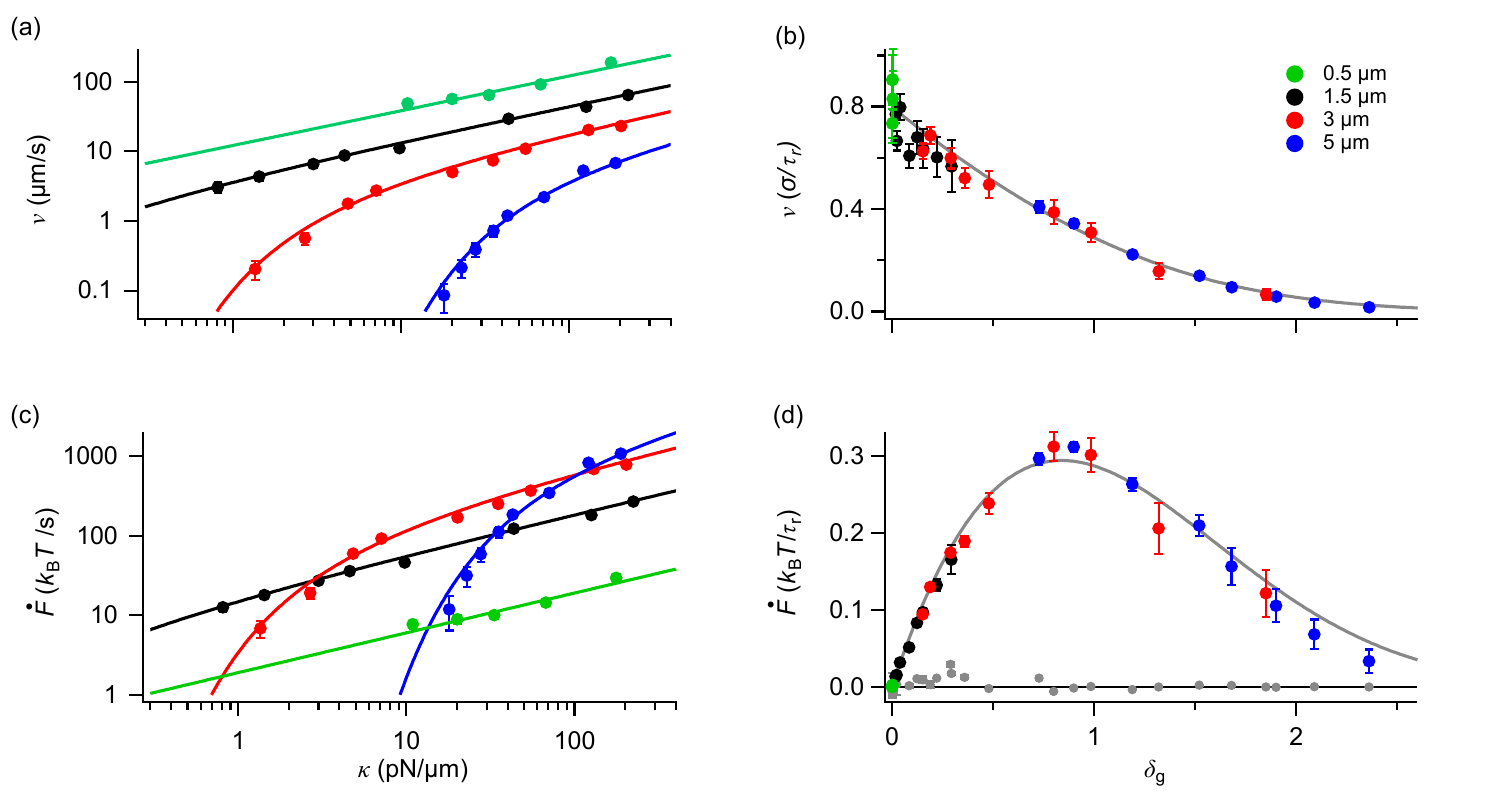}
    \caption{Dependence of engine performance on material parameters for a variety of bead diameters. (a) Directed velocity as a function of trap stiffness. (b) Scaled directed velocity as a function of scaled mass. (c) Output power as a function of trap stiffness. (d) Scaled output power as a function of scaled mass. Markers denote experimental data, with gray markers showing that the corresponding trap power remains small. The solid black curves in (b) and (d) represent semi-analytic calculations. Adapted from~\cite{Saha2021_Maximizing}.}
    \label{fig:pnasFigure4}
\end{figure}
Figure~\ref{fig:pnasFigure4} shows that the velocity $v$ and extracted power $\dot{F}$ each increase with increasing force constant $\kappa$, and each is proportional to $\sqrt{\kappa}$ as $\kappa \rightarrow \infty$, consistent with (\ref{vel_phys}) and (\ref{p_phys}). 

Engine performance depends on the scaled mass in a more complicated way. The directed velocity is maximized when $\delta_{\tn{g}} \rightarrow 0$: intuitively, as the gravitational force on the bead decreases, the bead fluctuates above the trap center more easily and ratchet events occur more frequently. In contrast, Fig.~\ref{fig:pnasFigure4}(d) shows that the extracted power is maximized at intermediate $\delta_{\tn{g}}$, balancing two competing effects: as $\delta_{\tn{g}}$ increases greater potential energy is stored per ratchet event, while the frequency of ratchet events decreases. 

In this experimental setup, we achieved power as high as 1,066 $k_{\tn{B}} T/\tn{s}$ and velocity as high as 190 \textmu m/s.  This highest velocity is $\sim$10x that of \textit{Escherichia coli} and comparable to the faster motile bacteria found in marine environments~\cite{barbara2003bacterial}, while the highest achieved power is comparable to that in molecular motors such as kinesin~\cite{ariga2018nonequilibrium}. With optimized material parameters, the maximum extracted power was greater than that of previous realizations of information engines, sometimes by orders of magnitude~\cite{toyabe2010experimental,lee2018experimentally, ribezzi2019large, Admon2018_Experimental, Paneru2018_Losless}.

\section{Nonzero trap work}\label{sec:non-zero}
So far, we have considered pure information engines that are powered solely by ratcheting mechanisms, not relying on conventional sources of fuel. More generally, engines could be powered by a mix of information and fuel, using feedback rules that allow nonzero input work~\cite{schmitt2015molecular}. We then quantify performance by the net output power, namely the rate of output free-energy storage minus the trap work rate.  We consider two broad categories of feedback schemes~\cite{lucero2021maximal}: \textit{unconstrained} schemes where all energy can be stored, and \textit{practical storage} schemes where only gravitational potential energy can be stored.

\subsection{Unconstrained scheme} 

When all energy can be stored (either as work via the trap or as a free-energy change via the gravitational potential), the feedback rule for trap updates should be capable of exploiting \textit{all} fluctuations---not just the `up' ones---and ratchet after each measurement. We therefore choose an affine feedback rule of the form 
\begin{equation} \label{update_uncon}
    \lambda_{n+1} = \lambda_{n} + \alpha (x_{n+1} - \lambda_{n}) + \psi \,.
\end{equation} 
While measurements are performed at intervals of 20 \textmu m (as before), we now assume that feedback is performed instantaneously following measurement, so that the update rule for $\lambda_{n+1}$ involves the bead position at time $n+1$, not $n$ as previously in (\ref{update_main}).

Optimizing the net output power over all feedback rules of the form (\ref{update_uncon}), we obtain optimal values $\alpha=1$ and $\psi=\delta_{\tn{g}}$ so that 
\begin{equation}
    \lambda_{n+1} - \delta_{\tn{g}} = x_{n+1} \,.
\end{equation}
This feedback rule moves the minimum $\lambda_{n+1} - \delta_{\tn{g}}$ of the \textit{total potential} to the last measured bead position $x_{n+1}$, thereby extracting the bead's total potential energy, similar to the feedback cooling in~\cite{lee2018experimentally}. The potential thus follows the bead at each time step, making the bead equally likely to fluctuate up or down in the interval between subsequent measurements. As a result, the bead has zero average directed velocity, so that $\dot{F} = 0$. In the continuous-sampling limit,
\begin{equation}
    P_{\tn{trap}} = -1 \,,
\label{eq:upper-bound}
\end{equation}
with corresponding net output power $P_{\tn{net}} = 1$, or $1 \, k_{\tn{B}} T/\tau_{\tn{r}}$ in dimensional units. This exceeds the net power when constrained to zero trap work and represents an upper bound on performance even when all energy can be extracted. Furthermore, this upper bound is independent of $\delta_{\tn{g}}$ and reflects a fundamental limit set by diffusion and the material parameters (trap stiffness and bead friction coefficient, entering through the relaxation time $\tau_{\tn{r}}$).

\subsection{Practical-storage scheme}
The best-performing unconstrained feedback rules lead to no net motion and all energy being transferred to the trap. While this does cool the bath, there is no practical way to store or use the extracted energy in the current setup. Practical schemes are thus currently limited to ones where the trap work is constrained to be nonnegative, which still permits `investing' trap work that may later be `repaid' by faster upward motion. We implement such feedback as follows: when the measurement $x_{n+1}$ exceeds the fluctuation threshold $\lambda_n + X\sub{T}$, we move the trap (instantaneously) according to \eqref{update_main} if the resulting trap work is nonnegative, otherwise we do not move the trap. By construction, this scheme enforces unidirectional energy flow from the trap to the bead. The zero-trap-work scheme of Section~\ref{sec:max_vel_power} is a special case of a nonnegative-trap-work scheme. 

\begin{figure}[tbh]
    \centering
    \includegraphics[width=0.7\linewidth]{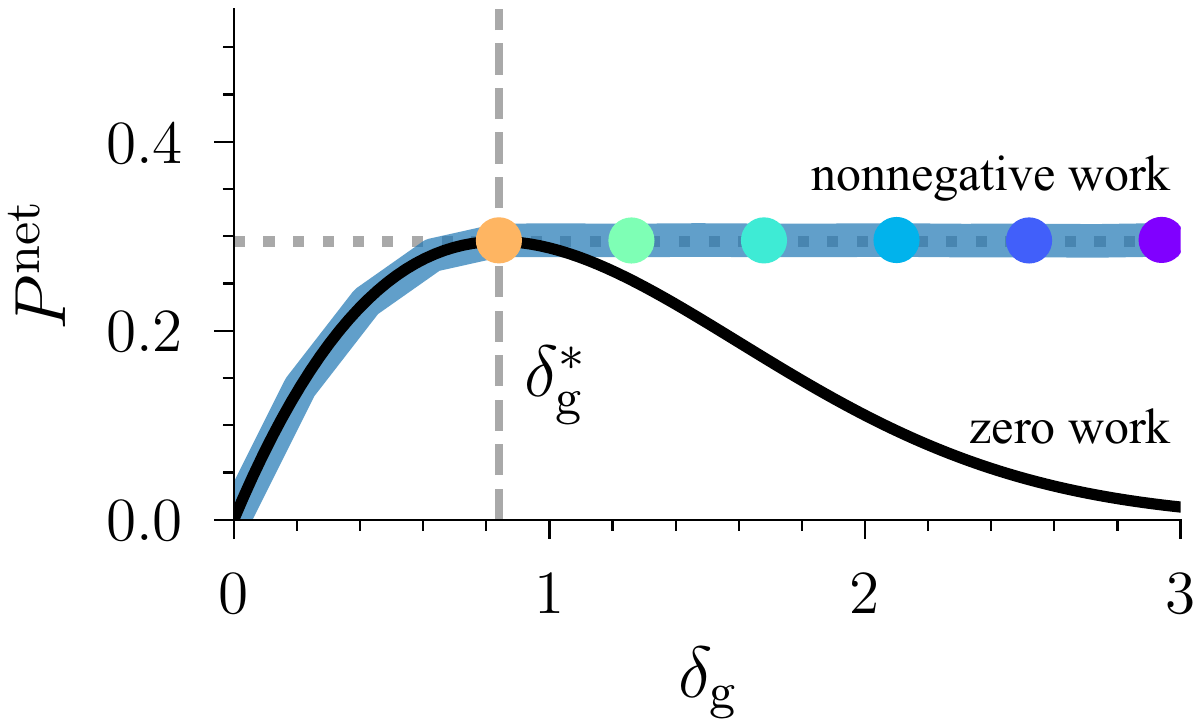}
    \caption{Maximum net output power when positive trap work is allowed (thick blue curve) or not (thin black curve).  Orange point denotes maximum net output power $P_{\tn{net}}$ for zero trap work at scaled mass $\delta_{\tn{g}}^*$.  Other colored points denote maximum $P_{\tn{net}}$ for various $\delta_{\tn{g}} > \delta_{\tn{g}}^*$. From~\cite{lucero2021maximal}.}
    \label{fig:luceroPREfig9a}
\end{figure}

We find a sharp transition in optimal strategy as $\delta\sub{g}$ crosses a critical value $\delta\sub{g}^*$ corresponding to the scaled mass that maximizes zero-trap-work output power (Fig.~\ref{fig:luceroPREfig9a}). For $\delta_{\tn{g}} < \delta\sub{g}^*$, the optimal nonnegative-trap-work scheme corresponds to the zero-trap work scheme. In contrast, for $\delta_{\tn{g}} > \delta_{\tn{g}}^*$, positive trap work is required to obtain the maximum net output power. In this range, this maximum net output power is independent of $\delta\sub{g}$ and equals the output power of the zero-trap-work scheme with optimal scaled mass $\delta\sub{g}^*$. 

Such a sharp change in optimal strategy as a function of system parameters is a common feature in control theory~\cite{bechhoefer2021control}. For small scaled mass ($\delta_{\tn{g}} < \delta_{\tn{g}}^*$), the cost of supplying additional trap work exceeds the resulting additional stored gravitational potential energy, which is proportional to $\delta_{\tn{g}}$. For heavier beads ($\delta_{\tn{g}} > \delta_{\tn{g}}^*$), where the effects of gravity are significant, a zero-trap-work engine has long periods without any ratchet events. Relaxing the zero-trap-work constraint permits a fluctuation threshold $X\sub{T} < 0$, thereby increasing the frequency of ratchet events and, consequently, the average directed velocity and output power. For heavier beads, these gains in output power compensate for the cost of input trap work. 

The maximal net output power for the nonnegative-trap-work scheme thus matches that of the zero-trap-work scheme, never exceeding $0.294~ k_{\tn{B}} T/ \tau_{\tn{r}}$ in dimensional units. Unconstrained engines that benefit from all fluctuations therefore extract more power than practical-storage engines that only benefit from a limited set of `up' fluctuations.

\section{Trajectory control}

We have shown that an information engine can rectify thermal fluctuations to create directed motion, thereby storing energy in a reservoir. This information engine does not control the bead's position to attain a particular value at a particular time; in fact, the variance of the bead position grows monotonically, diverging in the long-time limit. Here, following~\cite{saha2021trajectory}, we show that a pure information engine (operating with zero net trap work) can also control the bead's time-dependent position, exploiting fluctuations to follow a pre-specified trajectory. 

For simplicity, we only move the trap horizontally, so that we do not have to include the effects of gravity on the bead. The horizontal bead position evolves according to \eqref{langevin_scaled} with $\delta\sub{g}=0$. 

To make the bead follow a trajectory $r(t)$, we apply a zero-trap-work feedback rule that is similar to (\ref{update_main}), but with the following change: following each position measurement, if the trap force is in the desired direction (toward the current desired position) so that the bead will (on average) relax towards the desired position, then the trap does not move; if the force is in the `wrong' direction, then the trap moves to reverse the direction of the force. 

Given that the average directed velocity of a pure information engine is determined fundamentally by its material parameters and by the properties of the thermal fluctuations acting on it (Sec.~\ref{sec:max_vel_power}), a bead powered by an information engine cannot track a desired trajectory that changes faster than this same average velocity. Stated differently, the dependence on favorable thermal fluctuations limits the rate at which the engine can follow a desired trajectory.

\begin{figure}[tbh]
    \centering
    \includegraphics[width=\linewidth]{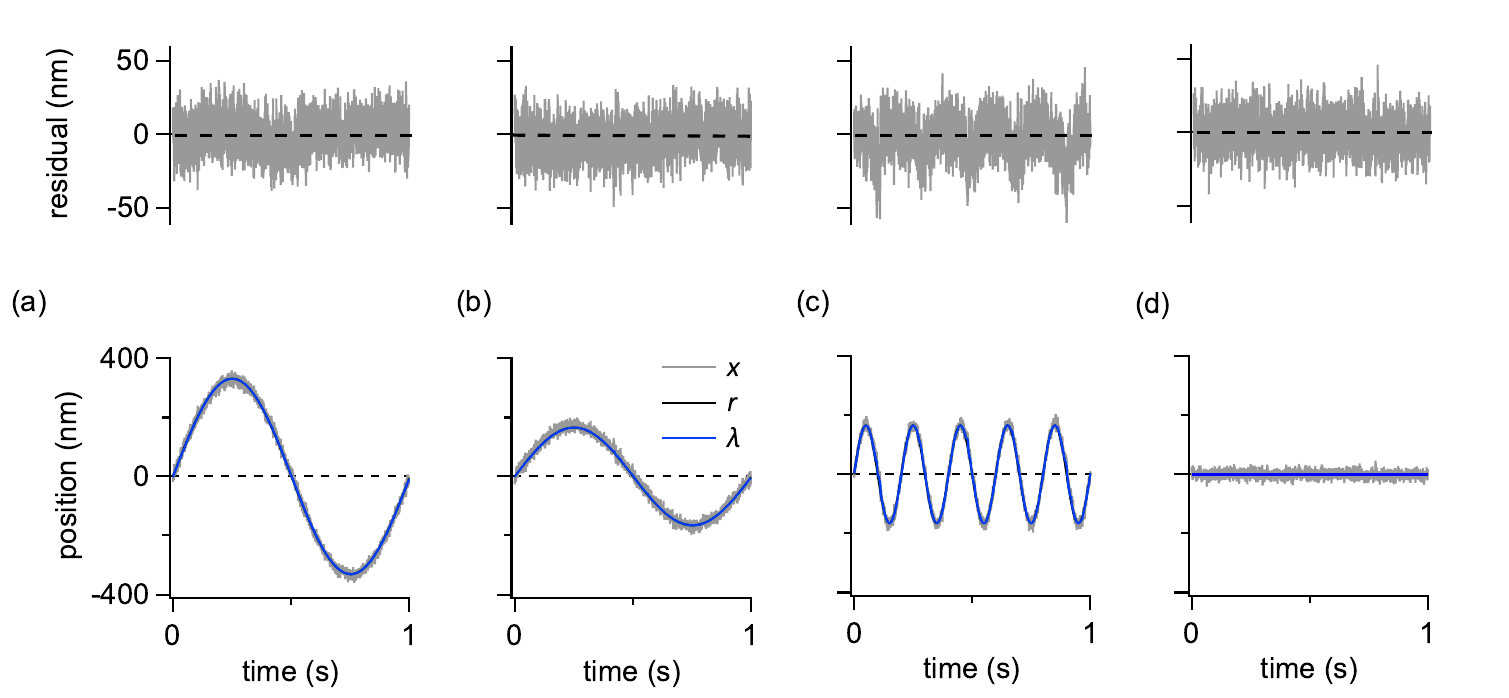}
    \caption{Controlling the bead trajectory to follow a sine wave. The bead position $x$ (gray) and trap center $\lambda$ (blue) for different desired trajectories $r$ (black), and the corresponding residual. The desired trajectories are (a) $2A\, \text{sin}(\omega t)$, (b) $A\, \text{sin} (\omega t)$, (c) $A \, \text{sin}(5\omega t)$, and (d) $0$, for $A=165$ nm and $\omega = 2\pi\times1$ Hz. From~\cite{saha2021trajectory}.}
    \label{fig:sine_wave}
\end{figure}

Figure~\ref{fig:sine_wave} illustrates tracking a sine wave, for different amplitudes and frequencies. We find that the tracking bandwidth is comparable to the maximum frequency that the engine can respond to, i.e., to the `corner frequency' of its power spectrum~\cite{saha2021trajectory}. The result implies that trajectory control is limited only by the engine's material parameters and not by the feedback algorithm employed. In our device, the spatiotemporal control of the bead was $\approx 10$ nm and 1~ms.

\section{Noisy measurement}

Previously, we assumed that the position measurements were error-free. But measurement noise is always present, degrading the information content of measurements and introducing the possibility of `incorrect' feedback decisions. Indeed, high levels of measurement noise can degrade performance so much that heat flow is reversed: instead of extracting energy from the heat bath, the information engine dissipates heat into the bath~\cite{paneru2020efficiency}. 

Previous attempts~\cite{paneru2020efficiency,taghvaei2021relation} to account for measurement noise have typically used `naive' feedback algorithms based directly on only the most recent measurement. Here, we contrast the performance of such a naive information engine with an information engine that also incorporates past measurements, optimally estimating the true position with a Bayesian filter~\cite{saha2022bayesian}. Our findings confirm theoretical studies~\cite{taghvaei2021relation,nakamura2021connection,horowitz2013imitating,rupprecht2020predictive} indicating that incorporating the information contained in past measurements can greatly improve the performance of an information engine.

\subsection{Naive protocol vs.\ Bayesian inference}
The experimental setup is largely the same as in Sec.~\ref{sec:materialsmethods}. The effect of measurement noise, practically implemented by reducing the intensity of the detection laser, results in measurements $y_n$ that differ from the true bead positions $x_n$ as 
\begin{equation}
    y_n = x_n + \sigma_{\tn{m}} \nu_n,
\end{equation}
for measurement noise strength $\sigma_{\tn{m}}$ and Gaussian random variable $\nu_n$ (independent of the thermal noise acting on the bead) with zero mean and covariance $\langle \nu_m \nu_n\rangle = \delta_{mn}$. Given that the bead's actual position fluctuates on a scale set by its equilibrium standard deviation $\sigma = \sqrt{k_{\tn{B}} T/\kappa}$, while the measurement noise fluctuates on a scale $\sigma_{\tn{m}}$, we define a signal-to-noise ratio $\tn{SNR} = \sigma/\sigma_{\tn{m}}$. 

We contrast the performance of a `naive' feedback protocol with that of a more sophisticated protocol which uses all past noisy measurements to infer the bead position. In the naive protocol, the trap is updated according to the zero-threshold rule (\ref{update_main}),
\begin{equation} 
\label{naive}
    \lambda_{n + 1} = \lambda_n + \alpha (y_n - \lambda_n) \, \Theta(y_n - \lambda_n), 
\end{equation}
with ratcheting based on the noisy observation $y_n$ that `naively' estimates the unknown true position $x_n$.

In the \textit{filtering} approach~\cite{bechhoefer2021control}, we first estimate the probability distribution for the current bead position $x_n$ based on all previous measurements $\{y_{n-1}, y_{n-2}, \ldots\}$. We then use the Fokker-Planck equation to predict the bead position at time step $n + 1$, based on the bead's linear dynamics. When the new measurement $y_n$ becomes available, the prediction is updated to a point estimate $\hat{x}_{n + 1}$ following the Bayes rule. This correction can be shown to minimize the variance between the actual and estimated position. No other unbiased estimator---linear or not---has a lower variance~\cite{kailath2000linear}. Our position estimate is thus the optimal estimate incorporating all past measurements. The feedback scheme is identical to \eqref{naive}, with $y_n \to \hat{x}_{n + 1}$, ratcheting when the predicted bead position $\hat{x}_{n+1}$ at the next time step exceeds the current trap center $\lambda_n$. The Bayesian prediction therefore accounts for the feedback delay $t_{\tn{s}}$, incorporating in its estimate the bead's linear dynamics between measurement and subsequent trap update.

\subsection{Results}

While the engine stores energy for any positive feedback gain, we restrict our attention to pure information engines, seeking positive values $\alpha^*$ in \eqref{naive} such that the trap work is zero (on average). To determine the respective values of feedback gain that ensure zero trap work (on average) for either the naive or Bayesian feedback schemes, we calculate the empirical trap work using \eqref{trap_work}, with $x$ replaced by the measurement outcome $y$, which can be shown to be an unbiased estimator of the true trap work for either scheme. 

\begin{figure}[tbh]
    \centering 
    \includegraphics[width = \linewidth]{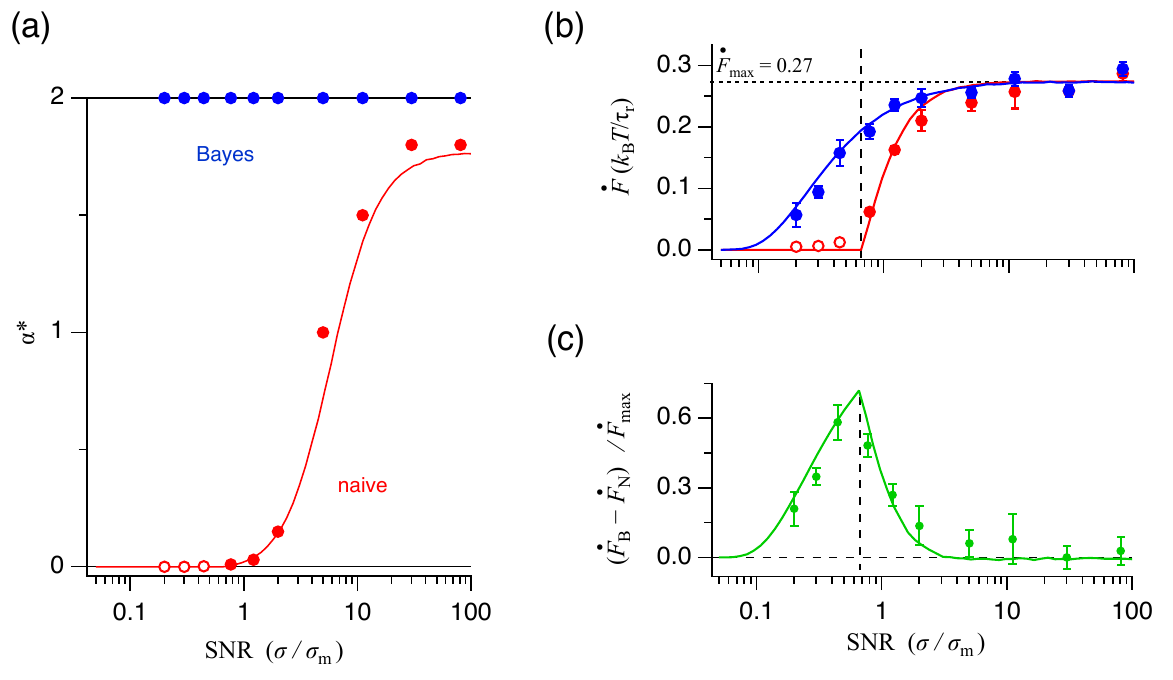}
    \caption{Information-engine performance. (a) Critical feedback gain $\alpha^*$ ensuring zero net work for naive (red) and Bayesian (blue) information engines. Solid red curve is from numerical simulation. (b) Output power of naive (red) and Bayesian (blue) information engines as 
    functions of SNR. Hollow red points denote SNRs for which $\alpha^* > 0$ could not be found. (c) Difference of output powers for the Bayesian and naive engines, scaled by the maximum power ($\dot{F}_{\tn{max}} = 0.27$). The difference peaks at SNR=SNR$_{\tn{c}}$~$\approx 0.7$ (vertical dashed lines). Points denote experimental means, solid curves numerical simulations. From~\cite{saha2022bayesian}.}
    \label{fig:BayesFigure2}
\end{figure}

Figure~\ref{fig:BayesFigure2}(a) illustrates that for the naive engine, below a critical $\tn{SNR}_{\rm c}$ ($= 0.7 \pm 0.1$) no positive $\alpha^*$ can be found that ensures zero trap work. Above that critical SNR, $\alpha^*$ increases and saturates at $\approx 1.5$ for large SNR. The vanishing of $\alpha^*$ below $\tn{SNR}_{\rm c}$ corresponds to a phase transition between a regime of successful operation (SNR $>$ SNR$_{c}$), where net output power $\dot{F} > 0$ can be achieved while simultaneously satisfying $P_{\tn{trap}} = 0$, and a `dud' regime (SNR $<$ SNR$_{c}$) where the engine cannot extract power without additional trap work~\cite{saha2022bayesian}.

In contrast, Fig.~\ref{fig:BayesFigure2}(a) shows that for all SNR the Bayesian engine has $\alpha^* \approx 2$, which is the feedback gain for an `ideal' pure information engine in the absence of measurement noise or feedback delay. This is because the Bayesian estimator accounts explicitly for both feedback delays and measurement noise. Thus, the Bayesian engine can extract energy without net trap work no matter how much measurement noise corrupts the signal, eliminating the phase transition between functional and non-functional regimes.

Fundamentally, the phase transition for naive feedback arises because a single noisy measurement is a biased estimate for the true bead position. First, the naive estimate neglects the feedback delay; second, the naive feedback treats all measurements as equally likely \emph{a priori}, even though fluctuations above the ratchet threshold are rarer than fluctuations below it. Since measurements above or below the true bead position are equally likely, for SNR<SNR$_\tn{c}$ most measurements above the threshold are `false positives' with the bead actually being below the trap center, so that ratcheting requires positive trap work; conversely, negative-trap-work ratchets are unlikely. This means there is no nonzero $\alpha^*$ for which the average trap work is zero. Because the Bayesian prediction is unbiased, energy extraction is possible for all SNR, and the phase transition is eliminated.

Figure~\ref{fig:BayesFigure2}(b) compares the output powers of the naive and Bayesian engines, showing that the Bayesian engine extracts more energy at all SNRs and illustrating the phase transition for the naive engine. Note that the two curves converge at high and low SNR: for high SNR$\gg 1$, the output powers of both naive and Bayesian engines saturate at the same finite value predicted from (\ref{p_phys}) in the absence of measurement noise, while for small measurement noise, the Bayesian estimate is approximately equal to the measurement, corrected for the feedback delay; this bias is small when the sampling time $t_{\tn{s}}$ is shorter than the bead relaxation time $\tau\sub{r}$ (20 vs.~1000 \textmu s). For low $\tn{SNR} \ll 1$, the variance of the measurement noise greatly exceeds the equilibrium variance of the bead position: negligible information can be extracted from measurements, whatever the estimator.

While the engines perform similarly at both low and high SNR, their performance differs markedly for intermediate $\tn{SNR} \lesssim 1$. Figure~\ref{fig:BayesFigure2}(c) shows that the difference in the engines' outputs is maximized at $\tn{SNR}\sub{c}$, coinciding with the phase transition in naive-engine performance. This increase in performance at intermediate SNR is a general feature of information processing by physical systems: not much is to be gained by sophisticated information processing when information is nearly perfect or when it is nearly useless, with the maximum benefit at intermediate noise levels~\cite{huang2018physically,bechhoefer2021control}.

\section{Nonequilibrium bath} 
As mentioned previously in Sec.~\ref{sec:materialsmethods}, the energetic costs associated with information processing equal or exceed an information engine's free-energy extraction when the engine and measurement apparatus are kept at the same temperature $T$; however, if the engine is in contact with a heat bath at a higher temperature, the resulting free-energy extraction can exceed the information-processing costs paid by the colder measurement apparatus~\cite{smoluchowski1912experimentell,feynman63the,parrondo1996criticism,still2020thermodynamic,bang2018experimental}.

Here, we investigate information-engine performance in the presence of both equilibrium fluctuations due to a heat bath at temperature $T$ as well as additional nonequilibrium fluctuations. Under these conditions, the information-to-work efficiency is limited only by the strength of the nonequilibrium forcing.

These results are of practical interest in scenarios where nonequilibrium fluctuations are generated by active media~\cite{ramaswamy2010mechanics,marchetti2013hydrodynamics,sokolov2010swimming,di2010bacterial,reichhardt2017ratchet,vizsnyiczai2017light,pietzonka2019autonomous,fodor2021active,speck2016stochastic,fodor2016far,mandal2017entropy,pietzonka2017entropy,dabelow2019irreversibility,caprini2019entropy,dabelow2021irreversibility} such as microswimmer suspensions~\cite{elgeti2015physics} and active Brownian particles~\cite{bechinger2016active} or in mesoscale biological systems, where active fluctuations in cellular environments serve as an additional source of fluctuations acting on machinery such as molecular motors~\cite{mizuno2007nonequilibrium,gallet2009power}. Our results indicate that in such contexts, information engines provide an advantage over conventional engines.

Moreover, understanding the operation of our information engine points to design principles for other energy-harvesting devices in environments where correlated non-thermal noise is important, such as sailboats~\cite{an2021autonomous} or wind turbines~\cite{porte2020wind, amano2017review}.

\subsection{System dynamics and nonequilibrium bath}

Following \cite{saha2023information}, we model the colloidal bead's motion in the optical trap as in (\ref{langevin_scaled}) but with an additional source of nonequilibrium fluctuations $\zeta$. This additional noise arises from the amplified, low-pass-filtered Johnson noise of a resistor that produces both electroosmotic and electrophoretic effects on the bead~\cite{cohen2007trapping}, and obeys the Langevin equation
\begin{equation} 
\label{neq_noise}
    f_{\tn{ne}}^{-1} \dot{\zeta}(t) = - \zeta(t) + \sqrt{2 D_{\tn{ne}}} \, \tilde{\xi}(t) \,,
\end{equation}
with $\tilde{\xi}$ a zero-mean Gaussian white noise having autocorrelation $\langle \tilde{\xi}(t) \tilde{\xi}(t') \rangle~=~\delta(t-t')$, independent of $\xi(t)$. The cutoff frequency $f_{\tn{ne}}$ is set by the filter frequency, and $D_{\tn{ne}}$ describes the strength of the electrokinetic forcing. A Langevin equation of the form (\ref{neq_noise}) produces colored Ornstein-Uhlenbeck noise~\cite{haunggi1994colored} with exponential autocorrelation 
\begin{equation} 
\label{autocorrelation_neq}
    \langle\zeta(t) \zeta(t') \rangle = D_{\tn{ne}} f_{\tn{ne}} \e^{-f_{\tn{ne}}|t - t'|}.
\end{equation}
By contrast, the thermal bath is assumed to equilibrate much faster than the bead relaxes in the trap, and is thus modeled as Gaussian white noise with no temporal correlations. 

Throughout this section, the trap dynamics are updated according to the feedback protocol in (\ref{update_main}), with the feedback gain $\alpha$ chosen to enforce zero trap work.

\subsection{Output power} \label{subsec:neq_output_power}

First, we investigate the dependence of the output power $\dot{F}$ on the parameters $f_{\tn{ne}}$ and $D_{\tn{ne}}$ that characterize the nonequilibrium fluctuations, and on the scaled mass $\delta_{\tn{g}}$. 
\begin{figure}
    \centering
    \includegraphics[width = 0.9\linewidth]{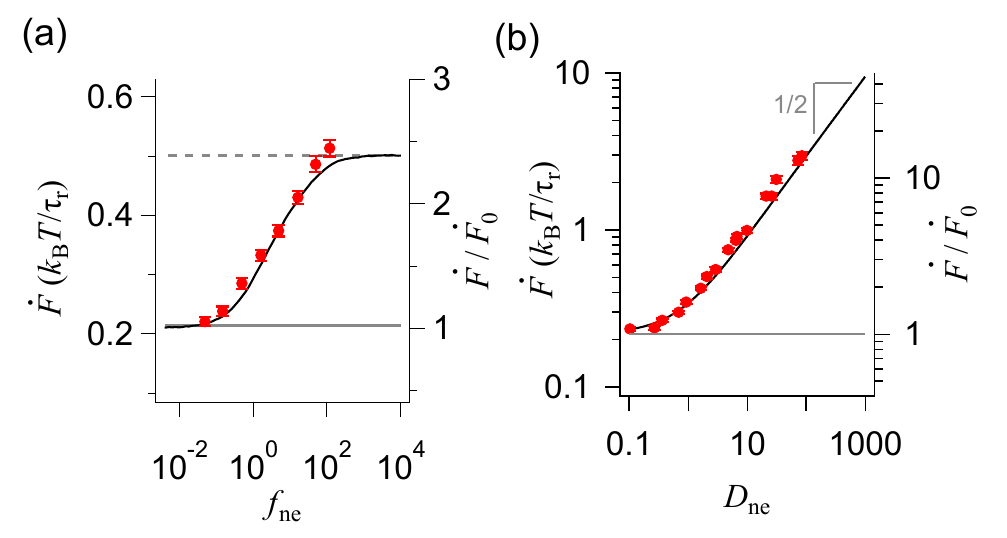}
    \caption{Nonequilibrium noise increases extracted power.  (a) Output power as a function of cutoff frequency $f_\tn{ne}$ of the nonequilibrium bath scaled by the trap cutoff frequency $f_\tn{c}$, at nonequilibrium noise strength $D_\tn{ne} = 3.0$ and scaled mass $\delta_\tn{g}=0.37$. Dashed line: high-$f_\tn{ne}$ limit. Right axis: ratio of nonequilibrium to equilibrium output powers. (b) Output power as a function of $D_\tn{ne}$ for high $f_{\tn{ne}}$ ($=118$). From~\cite{saha2023information}.}
    \label{fig:F_dep_Dne_fne}
\end{figure}
Figure~\ref{fig:F_dep_Dne_fne}(a) shows that the output power saturates for both $f_{\tn{ne}} \rightarrow 0$ and $f_{\tn{ne}} \rightarrow \infty$. For $f_\tn{ne} \rightarrow 0$, the variance of the nonequilibrium fluctuations vanishes, recovering the system studied previously in Sec.~\ref{sec:max_vel_power}, so that the output power equals (\ref{p_phys}). This is also the output power in the limit $D_{\tn{ne}} \to 0$ seen in Fig.~\ref{fig:F_dep_Dne_fne}(b). The output power increases monotonically with the cutoff frequency $f_{\tn{ne}}$, achieving more than half the maximum when the cutoff frequency of the nonequilibrium noise is only slightly greater than that of the engine. This reflects the fact that the majority of power supplied by the nonequilibrium bath is contained in the modes having frequency less than $f_{\tn{ne}}$. This contrasts with an equilibrium bath, which has modes with equal energies up to very high frequencies. In fact, the fraction of forced modes to equilibrium modes can be (for our system) of the order $10^{-10}$. When $f_{\tn{ne}}$ is 10--100$\times$ higher than the engine cutoff frequency, the nonequilibrium fluctuations become indistinguishable (to the engine) from those of Gaussian white noise, so that in this regime the nonequilibrium noise satisfies
\begin{equation}
    \zeta(t) = \sqrt{2 D_{\tn{ne}}}\, \tilde{\xi}(t).
\end{equation}
Thus, the bead can be viewed as being in contact with a thermal reservoir at an effective temperature $T_{\tn{ne}} = 1 + D_{\tn{ne}}$ in scaled units, so that the results of Sec.~\ref{sec:max_vel_power} can be applied here (taking into account the necessary rescalings required to incorporate this higher effective temperature). For our system, the nonequilibrium bath can have an effective temperature orders of magnitude higher than that of the thermal bath; because the majority of its power is contained in low-frequency modes, it can mimic the action of an extremely high-temperature thermal bath on our system (which is sensitive only to frequencies not much greater than $f_{\tn{c}}$), even though the total power contained in all its modes, if spread out equally across a large range of frequencies as in an equilibrium bath, would correspond to a low-temperature bath. Very high effective temperatures and correspondingly large power extraction is implicit in the functioning of sailboats, wind turbines~\cite{stavrakakis20122} and self-winding watches~\cite{parrondo1996criticism, watkins2016}, all of which must rectify nonequilibrium fluctuations. Furthermore, recent experiments show that ratchets driven by granular gases~\cite{eshuis2010experimental,joubaud2012fluctuation,gnoli2013brownian,gnoli2013nonequilibrium} can achieve effective temperatures $\sim 10^{17}$ K~\cite{rouyer2000velocity,feitosa2004fluidized,chastaing2017two}. 

In the white-noise regime ($f_{\tn{ne}} \gg f_{\tn{c}}$), the output power increases monotonically with increasing $D_{\tn{ne}}$. In fact, $\dot{F} \sim D_{\tn{ne}}^{1/2}$ when $f_{\tn{ne}}$ in Fig.~\ref{fig:F_dep_Dne_fne}(b), such that the additional power extractable in the presence of the nonequilibrium bath can be made arbitrarily large.

\subsection{Cost of feedback and information-to-energy efficiency}

The previous section showed that additional nonequilibrium fluctuations can drastically increase output power. Importantly, because the measurement and controller apparatus are kept at the temperature $T$ of the thermal bath, they are not directly influenced by the nonequilibrium noise strength $D_{\tn{ne}}$. This decoupling means that the extracted energy can exceed the fundamental energetic costs required for information processing and feedback, as we will show. 

The energetic cost of feedback must, at minimum, compensate for the entropy reduction due to the change in the controller (trap) position during feedback. More precisely, the additional work (normalized by $k_{\tn{B}}T$) to operate the controller must exceed the reduction, due to the controller's dynamics, in the conditional entropy~\cite{cover1999elements} $H[\Lambda|X]$ of the trap position given the bead position~\cite{ehrich2023energetic}. The \textit{information power}, measuring the rate at which energy is used to measure and erase information and update the trap, is thus bounded:
\begin{equation} 
\label{infopower}
    P_{\tn{info}} \geq \frac{H[\Lambda_{k-1}|X_k] - H[\Lambda_k|X_k]}{t_{\tn{s}}} \,.
\end{equation} 
This gives the minimum additional power that must be supplied to make the feedback procedure consistent with the second law of thermodynamics. The equipment used to operate the engine in our experiments exceeds the fundamental limit (\ref{infopower}) by orders of magnitude; nevertheless, studying the fundamental limits develops our understanding of the basic physical principles governing such systems. 

\begin{figure}[tbh]
    \centering
    \includegraphics[width = 0.9\linewidth]{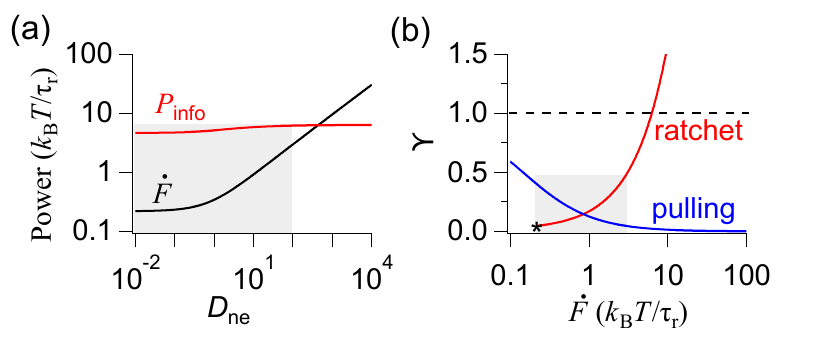}
    \caption{Numerical estimates of information-engine performance in the white-noise limit ($f_\tn{ne}=10^4$). (a) Output free-energy gain $\dot F$ and information power $P_\mathrm{info}$ (minimum power to perform measurement, erase information, and control the information engine) as functions of nonequilibrium noise strength $D_\tn{ne}$. (b) Performance $\Upsilon \equiv \dot{F}/P_{\tn{info}}$ as a function of $\dot F$, for information engine (red) or open-loop operation that simply moves the trap up at constant velocity (blue). Star indicates the equilibrium bath, $D_\tn{ne}=0$. Experimentally accessible regions are shaded. From~\cite{saha2023information}.} 
    \label{fig:Fout_vs_Pinfo}
\end{figure}

Figure~\ref{fig:Fout_vs_Pinfo}(a) compares the output power $\dot{F}$ with the information power $P_{\tn{info}}$, as functions of the nonequilibrium noise strength $D_{\tn{ne}}$, in the white-noise limit ($f_{\tn{ne}} \rightarrow \infty$). As noted in Sec.~\ref{subsec:neq_output_power}, for $D_{\tn{ne}} \gg 1$ the output power scales as $D_{\tn{ne}}^{1/2}$, whereas the information power saturates. In principle, this allows the extraction of orders of magnitude more power than that required to operate the information engine: Strong nonequilibrium fluctuations provide more opportunities for the trap to ratchet (and allow for larger ratchet events) without increasing the commensurate cost associated with information processing, which is set by the temperature $T$ of the controller apparatus. 

The output power divided by the minimal input power, $\Upsilon = \dot{F}/P_{\tn{info}}$, constitutes a natural performance measure. As described above, for sufficiently strong nonequilibrium noise the output power exceeds the operational costs, leading to $\Upsilon > 1$. This does not violate the second law, as $\Upsilon$ does not account for the power needed to generate the nonequilibrium fluctuations. 

Figure~\ref{fig:Fout_vs_Pinfo}(b) compares $\Upsilon$ with the efficiency $\dot{F}/\dot{W} < 1$ for a conventional engine, where the trap is moved up at constant velocity $v = \dot{F}/\delta_{\tn{g}}$ and with associated input power $\dot{W} = v^2 + \delta_{\tn{g}} v$. For small output power, the conventional engine outperforms the information engine, illustrating that when the difference between effective temperature $T_{\tn{ne}}$ and controller temperature $T$ is small, rectifying thermal fluctuations is an inefficient way to extract energy. As the output power grows (due to increasing nonequilibrium noise strength), the information engine can extract energy at lower cost than the conventional engine.

\section{Conclusions}

While previous studies on Maxwell demons have focused on demonstrating that their experimental realizations are consistent with the second law of thermodynamics and have explored their information-to-work efficiency, we have instead studied the performance limits of an information engine. Our engine, an optically trapped bead, converts energy from favorable fluctuations of a heat bath into directed motion, thereby storing gravitational potential energy. As illustrated in Fig.~\ref{fig:infoVconventional}b, we employ feedback protocols that store this energy without any direct external work, isolating the effect of information on the engine's performance.

We found that performance, judged by the average directed velocity and power output, is limited only by the system's material parameters, such as the bead's scaled mass and the trap stiffness. The same is true for the engine's ability to track a desired signal, which is limited only by the trap's dynamical response.

Allowing for nonzero trap work, unconstrained strategies drastically outperform strategies where trap work cannot be stored. For engines that can store all available energy, the optimal strategy is to move the trap to extract all available potential energy: all fluctuations are exploited. `Practical storage' schemes benefit only from `up' fluctuations and thus produce significantly less output power. Finding a design for a colloidal engine that mimics more closely the Szilard-engine thought experiment and stores energy from all measurement outcomes---`down' as well as `up' fluctuations---could double performance.

In practical scenarios, information and, consequently, performance are degraded by measurement noise. Strikingly, as SNR decreases, we found a phase transition at a critical SNR, above which the engine extracts energy without external work and below which the engine cannot function without external work. However, using a feedback protocol that exploits all past measurements and optimally predicts the bead's position using Bayesian inference, an information engine can extract energy at all SNRs without requiring external work. The resulting improved performance at low SNR can reduce the thermodynamic costs of the information processing needed to operate the engine and also benefit experimental investigations of motor mechanisms that use fluorescent probes~\cite{veigel2011moving}, where lower light intensities can extend observation times.

The colloidal engine we investigated could be improved: For example, although we have explored a large class of feedback algorithms, we have not proven that any of them is optimal among all possible algorithms. In particular, in nonequilibrium environments, a feedback algorithm may profitably exploit known temporal correlations of fluctuations~\cite{malgaretti2022szilard,cocconi2024on}, much in the way that velocity feedback improves control in underdamped systems~\cite{kim2004entropy,bechhoefer2021control}.

Also, in the information engine discussed here, measurement and information processing take place external to the colloidal system. Realizing an autonomous version~\cite{koski2015chip} of this colloidal information engine, where engine and measuring device are both implemented via the colloidal physics, would better illuminate the information and energy flows and increase our understanding of the fundamental performance limits of information-fueled engines.

A particularly striking aspect of information engines is that nonequilibrium fluctuations that act on the bead but not the controller can improve performance by orders of magnitude. The huge effective temperatures induced by nonequilibrium fluctuations suggest the potential for drastic efficiency increases over machines operating solely between two thermal reservoirs. Active cyclically operating heat engines indeed show significant performance increases~\cite{datta2022second,krishnamurthy2016micrometre,zakine2017stochastic,lee2020brownian,saha2019stochastic,kumari2020stochastic,ekeh2020thermodynamic,holubec2020active,gronchi2021optimization,roy2021tuning}. Furthermore, since the controller dynamics are decoupled from these additional fluctuations, the output power can exceed the information-processing costs needed to operate it. This illustrates a possible advantage of information engines in nonequilibrium environments, such as in the biological cell. And if one passes from the behavior of isolated information engines to multiple information engines that interact with each other (such as exist inside cells), the result can show qualitatively new behavior. For example, VanSaders and Vitelli have recently shown that interacting autonomous information engines can develop spontaneous, lifelike collective motions~\cite{vanSaders2023informational}.

At large scales, organisms are known to take advantage of nonequilibrium motion: Birds soar on thermal plumes~\cite{cone1962the,reddy2016learning}; plankton surf on turbulent flows~\cite{monthiller2022surfing}. And, at molecular scales, \emph{in vitro} nonequilibrium fluctuations do accelerate the operation of kinesin, accounting for as much as 75\% of its velocity~\cite{ariga2021noise}. But \textit{in vivo} it is an open question whether these and other molecular motors are powered (in part) by such fluctuations. If nonequilibrium fluctuations do contribute significantly as a power source to molecular motors---if nonequilibrium forces are comparable to those produced by ATP hydrolysis~\cite{NancyAR}---then the mechanisms by which motors sense and exploit these fluctuations and the consequences of any collective effects would be important to understand.

\section*{Acknowledgments}

This review focuses on a research program published in several papers, for which we acknowledge the other authors:
Tushar K.\ Saha, Jannik Ehrich, Joseph N.\ E.\ Lucero, Susanne Still, and Momcilo Gavrilov. We also thank Tushar K.\ Saha for help in editing figures and for helpful conversations.

\section*{Disclosure statement}

No potential conflict of interest was reported by the authors.

\section*{Funding}

This research was supported by grant FQXi-IAF19-02 from the Foundational Questions Institute Fund, a donor-advised fund of the Silicon Valley Community Foundation. Additional support was from the Natural Sciences and Engineering Research Council of Canada (NSERC) Discovery Grants (JB and RGPIN-2020-04950 for DAS) and a Tier-II Canada Research Chair CRC-2020-00098 (DAS).

\section*{ORCID}

Johan du Buisson \url{http://orcid.org/0000-0003-4241-4554} \\
David A.\ Sivak \url{https://orcid.org/0000-0003-4815-4722} \\
John Bechhoefer \url{https://orcid.org/0000-0003-2568-5491} \\

\bibliography{performance.bib}

\end{document}